# Plasmonic Toroidal Metamolecules Assembled by DNA Origami

Maximilian J. Urban,[1,†] Palash K. Dutta,[1,‡] Pengfei Wang,[1,‡] Xiaoyang Duan,[†] Xibo Shen,[†] Baoquan Ding,[*,∥] Yonggang Ke,[*,‡] Na Liu,[*,†,§]

[†] Max Planck Institute for Intelligent Systems, Heisenbergstrasse 3, D-70569 Stuttgart, Germany

[‡] Wallace H. Coulter Department of Biomedical Engineering, Georgia Institute of Technology and Emory University, Atlanta, Georgia 30322, United States

[∥] CAS Key Laboratory of Nanosystems and Hierarchical Fabrication, CAS Center for Excellence in Nanoscience, National Center for Nanoscience and Technology, Beijing 100190, China

[§] Kirchhoff Institute for Physics, University of Heidelberg, Im Neuenheimer Feld 227, D-69120, Heidelberg, Germany

*Supporting Information Placeholder*

**ABSTRACT:** We demonstrate hierarchical assembly of plasmonic toroidal metamolecules, which exhibit tailored optical activity in the visible spectral range. Each metamolecule consists of four identical origami-templated helical building blocks. Such toroidal metamolecules show stronger chiroptical response than monomers and dimers of the helical building blocks. Enantiomers of the plasmonic structures yield opposite circular dichroism spectra. The experimental results agree well with the theoretical simulations. We also demonstrate that given the circular symmetry of the structures, distinct chiroptical response along their axial orientation can be uncovered via simple spin-coating of the metamolecules on substrates. Our work provides a new strategy to create plasmonic chiral platforms with sophisticated nanoscale architectures for potential applications such as chiral sensing using chemically-based assembly systems.

Realization of three-dimensional (3D) plasmonic nanoarchitectures exemplifies one of the key challenges in nanotechnology.[1] In general, there are two basic types of nanofabrication strategies: top-down and bottom-up. Top-down approaches such as electron-beam lithography and focused-ion beam etching offer custom patterns with relatively low throughput and resolution.[2] The fabricated 3D plasmonic nanostructures are generally fashioned in layers and spaced with dielectrics. This posts many challenges for downstream applications. In contrast, bottom-up self-assembly approaches,[3] which often involve molecular recognition characters, offer large-scale fabrication, biochemical compatibility, and many other benefits. Among a variety of materials for self-assembly, DNA represents one of the most attractive building blocks largely due to its unprecedented programmability. DNA has been used to construct increasingly complex structures[4] and to pattern nanoparticles.[5] In particular, the DNA origami[4] technique allows for creation of fully addressable DNA templates with custom-designed shapes and sizes. DNA origami enables rational 3D organization of metal nanoparticles with nanometer precision, allowing for engineering complex plasmonic architectures with tailored optical response.

Recently, plasmonic chiral nanomaterials have gained increasingly attention due to its potential applications in nanophotonics, ultrasensitive sensing, and *etc*.[6] Natural chiral molecules such as proteins and DNA only exhibit weak optical activity in the UV range. Plasmonic assemblies composed of metal nanoparticles that are arranged in chiral configurations offer an elegant way to achieving strong optical activity in the visible spectral range.[6] For example, in the seminal work of Liedl *et al*, plasmonic helices were realized by assembling gold nanoparticles (AuNPs) on origami bundles in a staircase fashion[7] Nevertheless, previous work has mostly focused on studies of fundamental plasmonic chiral building blocks (*e.g.* helices[7] and tetramers[8]), and averaged chiroptical response over all structural orientations, because the assemblies were randomly oriented in aqueous solutions.

In this communication, we demonstrate the first experimental realization of plasmonic toroidal metamolecules through hierarchical assembly of helical building blocks templated by DNA origami. Each structure consists of four plasmonic helical monomers arranged in a toroidal geometry.[9] Systematic investigations on different experimental conditions are carried out to achieve high-quality plasmonic structures for tailored optical functionalities. The assembled metamolecules with designated handedness exhibit pronounced optical activity in the visible spectral range. The experimental optical spectra agree well with the theoretical predictions. We also demonstrate that given the unique circular symmetry, distinct chiroptical response along the axial orientation of the toroidal metamolecules can be revealed by simple spin-coating of the structures on substrates. Such orientation self-alignment neither requires surface functionalization nor introduces birefringence effects, which often arise from symmetry breaking. This is of great importance for achieving reliable chiroptical results on the single structure level using dark-field spectroscopy as well as very practical for development of polarization conversion devices using chemically-based assembly systems. In addition, the axial chiroptical response can be enhanced through silver (Ag) coating of the AuNP-based structures. Elemental mapping with energy-filtered transmission electron microscopy (TEM) clearly demonstrates the formation of Au core/Ag shell nanoparticles in a toroidal geometry.

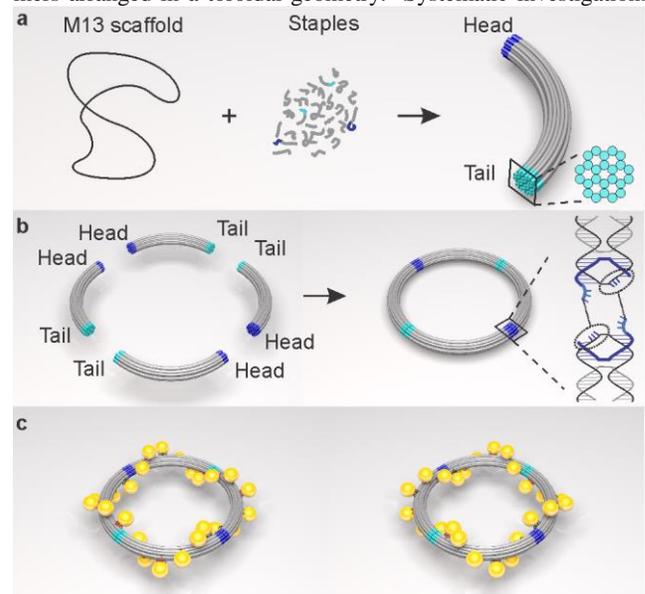

Figure 1. (a) A curved DNA origami monomer that comprises 24 DNA helices in a bundle is formed by a scaffold strand and numerous staple strands. (b) Four origami monomers form an origami ring through head-head and tail-tail connectors. (c) Schematic of the left-handed (LH) and right-handed (RH) 3D plasmonic toroidal metamolecules. Each metamolecule has a toroidal diameter of 120 nm and consists of 24 AuNPs (13 nm).

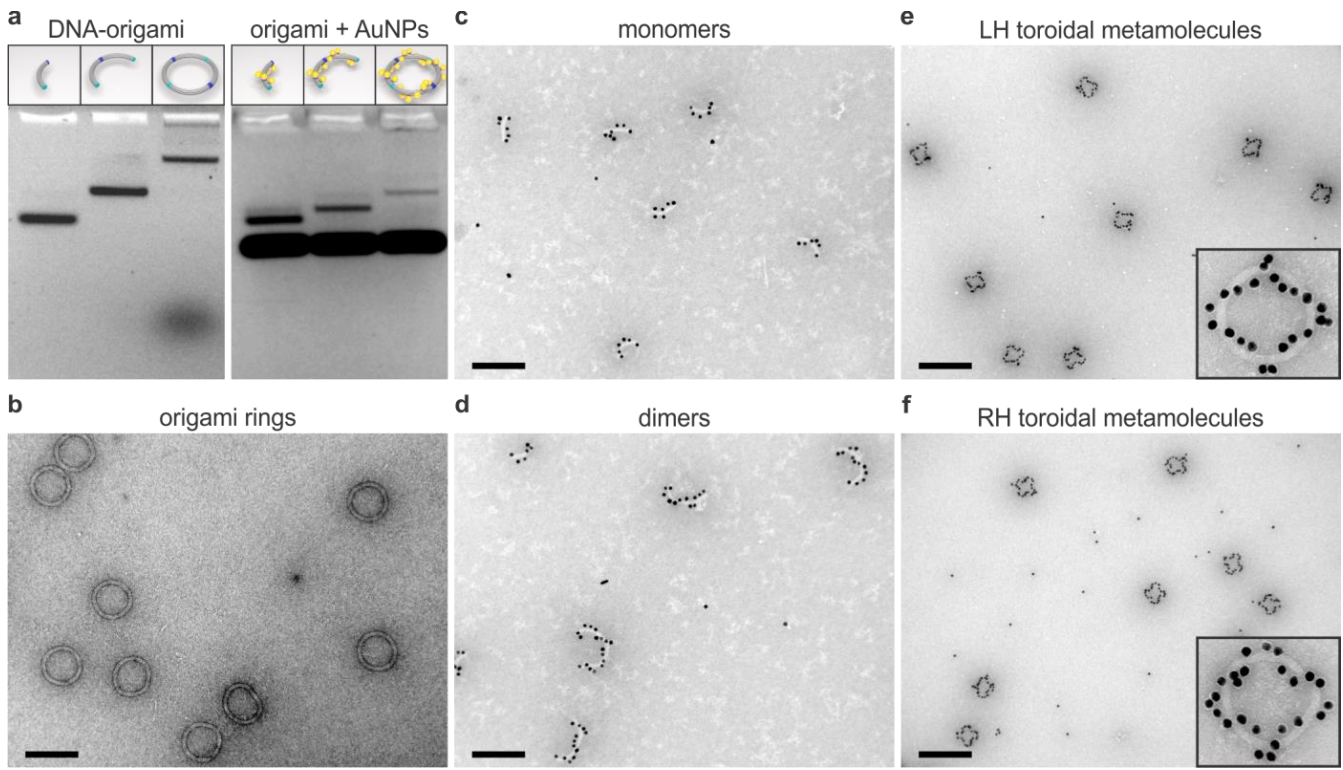

Figure 2. (a) Left: Agarose gel electrophoresis images of the origami monomers, dimers, and rings. Right: Agarose gel electrophoresis images of the plasmonic LH helical monomers, dimers, and toroidal metamolecules. (b) TEM image of the origami rings (scale bar 200 nm). (c) TEM image of the LH plasmonic helical monomers (scale bar 200 nm). (d) TEM image of the LH plasmonic helical dimers (scale bar 200 nm). TEM images of the LH (e) and RH (f) plasmonic toroidal metamolecules (scale bar 400 nm). The frame size of each inset image is 200 nm.

The plasmonic toroidal metamolecules are assembled utilizing DNA origami technique, as illustrated in Fig. 1. A long single-stranded DNA scaffold (M13 bacteriophage DNA) is folded by ~180 short staple strands through hybridization to form an origami monomer (Figs. 1a, S1, and S2; also see supporting information for more design details). Each monomer contains 24 curved DNA helices bundled in a honeycomb lattice.[10] Four monomers are connected together using 12 head-head and 12 tail-tail connector-strands to form a complete origami ring of ~120 nm in terms of inner-diameter (Figs. 1b and S3). Half-ring structures, *i.e.*, dimers, can be achieved by linking two monomers using either the head-head or tail-tail connectors. To assemble AuNPs on origami, six binding sites arranged either in left-handed (LH) or right-handed (RH) fashion are extended from each monomer. Each binding site contains three capture strands of the same sequence. Overall, 24 binding sites are distributed evenly along one complete origami ring. To ensure that each AuNP binds only to one site, two sets of capture strands with different sequences are used for the binding sites on the monomers in an alternating fashion. AuNPs (13 nm in diameter) functionalized with complementary DNA strands are assembled onto the origami ring to form a LH or RH plasmonic toroidal metamolecule (Fig. 1c).

Previously, Shih *et al.* have demonstrated a gear-like origami structure of ~100 nm in diameter using four 18-helix homo-monomers but the yield was lower than 10%.[11] To ensure a high yield of our origami rings, which is critical to create high-quality plasmonic structures for strong chiroptical response, careful optimization of the ring design and assembly protocol have been carried out. First, we have tested two versions (V1 and V2) of monomer designs with different degrees of curvatures (Figs. S1-S7). Mean angles of the V1 and V2 monomers are 99±9° and 79±12°, respectively (Fig. S7). The V1 monomers show higher yield for forming origami rings, while the V2 monomers give rise to mainly larger concatemers. Hence, subsequent optimization has been carried out using the V1 origami.

Second, the influence from the sticky-end strength of the connectors on the hierarchical assembly[12] of the origami rings is investigated. Each connector comprises two segments. One segment is anchored on the edge of one monomer and the other segment that contains 12 sticky-ends (either 1-base or 3-base long) is anchored on the edge of another monomer. Systematic studies using gel electrophoresis and TEM clearly show that connectors with 3-base long sticky-ends give rise to higher yield than those with 1-base long sticky-ends (Fig. S8a). Stronger sticky-ends with higher binding forces may minimize non-specific binding and thus facilitate the formation of the origami rings.

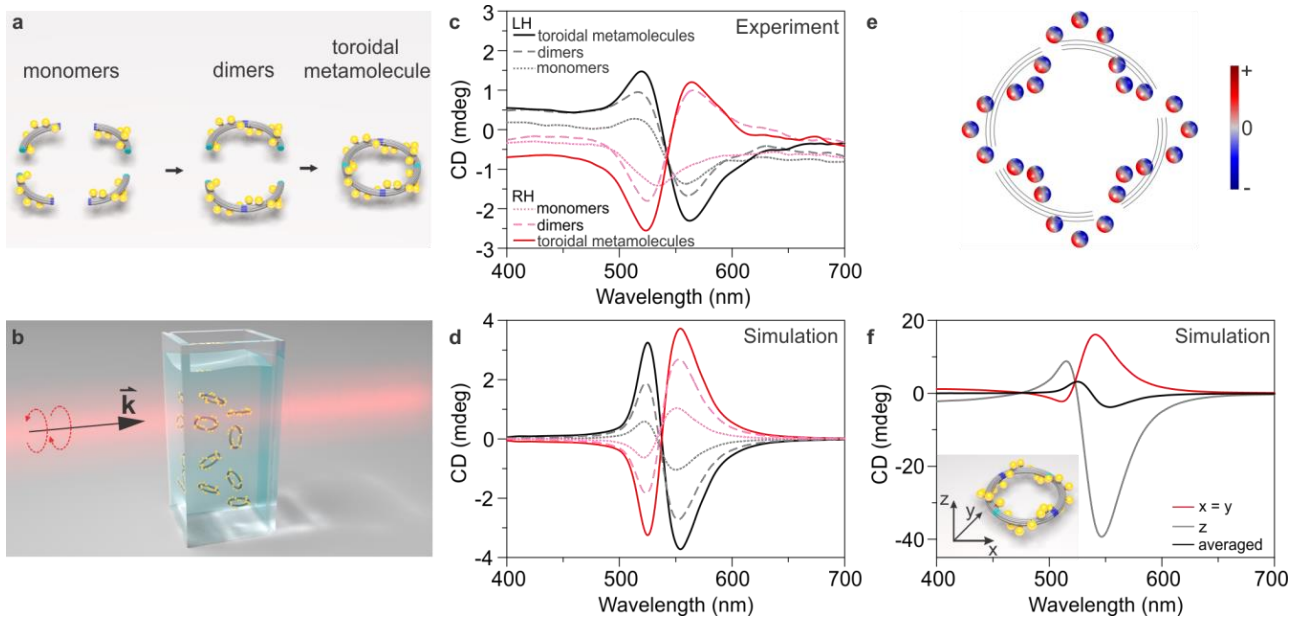

Figure 3. (a) Stepwise formation of the plasmonic toroidal metamolecules. (b) Schematic for the CD measurement of the plasmonic nanostructures that are dispersed in water with random orientations. Measured (c) and simulated (d) CD spectra of the plasmonic helical monomers, dimers, and toroidal metamolecules. (e) Charge distribution ($E_z$) of the plasmonic structure at resonance. (f) CD components of the LH plasmonic toroidal metamolecules along different incident directions of the circularly polarized light. The averaged CD response is presented by the black curve.

A comparison between the rings formed either directly or in a stepwise manner from purified monomers and the rings formed from purified dimers shows no substantial difference in yield (Fig. S8b). At last, the effects of the dimer concentration, assembly time, and connector-strand concentration on the assembly yield of the origami rings are examined (Figs. S9 and S10). Experimental results show that lower dimer concentrations (0.4 and 0.2 nM) lead to significantly improved yield (~49% and ~67%, respectively) of the origami rings, when compared to 32% from a higher dimer concentration of 1 nM. The incorrect assemblies are mostly aggregated spiral-like structures as observed from gel electrophoresis and TEM images (Fig. S9). Lower assembly concentrations are associated with slower assembly rates and lower possibilities of aggregation, which could both contribute to higher yield of the origami rings. In another assay experiment on various assembly times, many incomplete rings are observed, when the incubation time is less than 6 hours (Fig. S10a). Structure analysis using gel electrophoresis shows that a minimal 50-fold excess of the connectors is required to achieve high yield of the origami rings (Fig. S10b). Utilizing the optimized design and assembly protocol, DNA-origami monomers, dimers, and rings are successfully assembled with high yield, as confirmed by gel electrophoresis (Fig. 2a, left panel) and TEM imaging (Fig. 2b). The inner-diameter of the rings is measured to be 120.0±3.9 nm based TEM images (Fig. S11). Subsequently, plasmonic helical monomers, dimers, as well as toroidal metamolecules in LH and RH geometries are assembled via attaching AuNPs onto the origami precursors. AuNPs functionalized with single-stranded DNA are mixed with the corresponding purified origami at a 10-fold excess of particles per binding site. Both gel electrophoresis (Fig. 2a, right panel) and TEM images (Figs. 2c-2f) confirm the successful assembly of the plasmonic nanostructures (Fig. S12). The plasmonic toroidal structures were purified by gel electrophoresis prior to TEM imaging (Fig 2e, f). Particle numbers fewer than 24 are observed from the individual plasmonic structures. This is likely due to steric hindrance between the AuNPs as well as due to the inner-capture strands that are less accessible than the outer ones (Fig. S13).

When interacting with light, plasmons excited in the AuNPs can be collectively coupled.[6] The helical arrangement of the AuNPs imprints a "corkscrew" character to the collective plasmons, introducing a twist of specific handedness into the propagation direction of light, which leads to different absorption in response to LH and RH circularly polarized light, i.e., circular dichroism (CD).

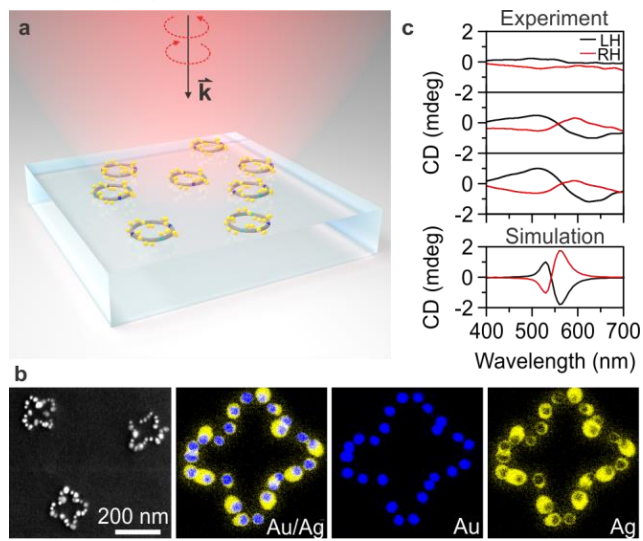

Figure 4. (a) Schematic for the axial CD component measurement of the plasmonic toroidal metamolecules. Circularly polarized light is perpendicularly incident on the structures. (b) Left: SEM image of the toroidal metamolecules after Ag enhancement. Right: Energy-filtered TEM images of the Ag-enhanced metamolecules. Au and Ag are represented by the blue and yellow colors, respectively. Frame size: 205 nm (c) Measured and simulated CD spectra of the LH and RH samples after one (top), three (middle), and ten (bottom) spin-coating cycles of the respective structures on the glass substrates. The structure density after ten spin coating cycles increased to approximately ~0.15 ring/μm², which is also utilized in the simulations.

In order to characterize the chiroptical response of the plasmonic helical monomers, dimers, and toroidal metamolecules, CD

measurements have been carried out using a Jasco-1500 CD spectrometer. To keep the AuNP concentration in all measurements consistent, the CD spectra are measured from the same solution by first addition of the head-head connectors into monomers to form dimers and subsequent addition of the tail-tail connectors to form toroidal metamolecules (Figs. 3a and 3b). The LH monomers, dimers, and toroidal metamolecules exhibit characteristic peak-to-dip line shapes with increasing amplitudes, while their RH counterparts display mirrored spectra correspondingly. The plasmonic toroidal metamolecules show the most pronounced CD spectra, when compared to the constituent plasmonic helical monomers and dimers (Figs. 3c and S14). The experimental results agree well with the simulated results (Fig. 3d). The charge distribution at resonance is presented in Fig. 3e, in which the dipole coupling nature between the neighboring AuNPs can be clearly identified. In the aqueous solution, the plasmonic metamolecules are randomly oriented and illuminated by light of fixed incident direction (Fig. 3b). In simulations, this is equivalent to averaging over the CD components along all possible incident directions. Fig. 3f presents the simulated CD components along different directions, as well as the averaged CD response of the LH plasmonic toroidal metamolecules.

Next, the CD response along the axial orientation of the toroidal metamolecules is explored by spin-coating the respective LH and RH structures on glass substrates after silver enhancement. The incident direction of light is perpendicular to the substrate and therefore the axial CD component of the structures becomes accessible (Fig. 4a). Silver enhancement is utilized in order to achieve enhanced CD signals using a relatively small amount of the structures (Fig. S15). The deposition of Ag shells on the AuNPs in the assemblies is verified by scanning electron microscopy (SEM) imaging and elemental-mapping imaging using energy-filtered TEM (Fig. 4b). A 1.5 μL of the LH or RH sample is first spin-coated on a glass substrate at 4000 rpm, which yields a very weak CD response (Fig. 4c, top panel), owing to a low structure density of $< 0.1$ rings/μm$^2$ (Fig. S16). Through repetitive spin-coatings, the structure density is eventually increased to ~0.15 ring/μm$^2$ (bottom panel). As a result, the CD response becomes increasingly stronger. The LH and RH samples exhibit bisignate spectral profiles, which are nearly mirrored (Fig. 4c). The experimental spectra show broader spectral profiles when compared to the simulated spectra (Fig. 4c) because of inhomogeneous broadening mainly resulting from the Ag shell thickness variations over different AuNPs. In contrast, a uniform Ag shell (5 nm) over different AuNPs is utilized in the simulations. The spin coating and drying process may also cause distortion to the toroidal geometry. The elimination of birefringence effects is confirmed by measuring all the samples from both the front and back incident directions (Fig. S17). Simulations on toroidal dimer structures and fused toroidal structures can be found in Figs. S18 and S19, respectively.

In conclusion, we have demonstrated DNA origami-templated plasmonic toroidal metamolecules, which yield tailored optical activity in the visible spectral range. These plasmonic toroidal metamolecules show stronger chiroptical response than the constituent plasmonic chiral building blocks such as helical monomers and dimers. These toroidal metamolecules can be dispersed on solid substrates for generating chiroptical response along their axial orientation. This might offer a new pathway to create plasmonic platforms with tunable CD for enantiomer sensing. The CD intensity dependence on the structure density could also be useful for development of polarization conversion devices, which allow for specific polarizations of light with controlled intensities.

## ASSOCIATED CONTENT

## Supporting Information

Detailed experimental procedures, origami design, DNA sequences, gel and TEM images, and structure density analysis after spincoating. This material is available free of charge via the Internet at http://pubs.acs.org.

## AUTHOR INFORMATION

### Corresponding Authors

*laura.liu@is.mpg.de;*yonggang.ke@emory.edu;
*dingbq@nanoctr.cn

### Author Contributions

1: M.J.U.; P.K.D. and P.W. contributed equally to this work.

### Notes

The authors declare no competing financial interests.

## ACKNOWLEDGMENT

We thank K. Hahn for assistance with elemental mapping and HR-TEM microscopy. TEM data was collected at the Stuttgart Center for Electron Microscopy (StEM). We thank Thomas Weiss and Anna Asenjo-Garcia for initial calculations. This research was supported by the Sofja Kovalevskaja grant from the Alexander von Humboldt-Foundation, the Marie Curie CIG grant, and the European Research Council (ERC Dynamic Nano) grant. We also thank the support from the Wallace H. Coulter Department of Biomedical Engineering Faculty Startup Grant and a Winship Cancer Institute Billi and Bernie Marcus Research Award.

M.; Zhang, C.; Ribbe, A. E.; Jiang, W.; Mao, C. *Nature* **2008,** *452*, 198- 201.

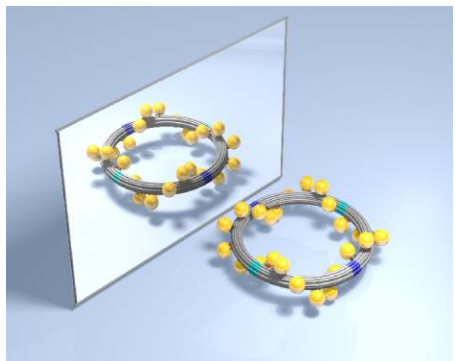